\documentclass[aps,prl,preprint,unsortedaddress,showkeys,nofootinbib]{revtex4-1}

\usepackage{amssymb,amsmath,amsfonts}

\usepackage{graphics}
\usepackage{graphicx}
\usepackage{epstopdf}
\usepackage[pdftex]{hyperref}

\usepackage{bm}
\usepackage{subfigure}

\usepackage{color}
\usepackage{setspace}
\usepackage{multirow}

\usepackage[ruled,vlined,linesnumbered]{algorithm2e}

\usepackage{color, colortbl}
\definecolor{Yellow}{rgb}{1,1,0}
\definecolor{Grey}{rgb}{.87,.87,.87}
\definecolor{Purple}{rgb}{.8,.0,1.0}
\definecolor{Crimson}{rgb}{.86,.08,.23}

\begin{document}

\title{A General Optimization Technique for High Quality Community Detection in Complex Networks}

\author{Stanislav Sobolevsky
\footnote{To whom correspondence should be
addressed: stanly@mit.edu}}
\author{Riccardo Campari}
\affiliation{SENSEable City Laboratory,
Massachusetts Institute of Technology, 77 Massachusetts Avenue,
Cambridge, MA 02139, USA}
\author{Alexander Belyi}
\affiliation{Belarusian State University, 4 Nezavisimosti Avenue, Minsk, Belarus}
\affiliation{SENSEable City Laboratory,
Massachusetts Institute of Technology, 77 Massachusetts Avenue,
Cambridge, MA 02139, USA}
\author{Carlo Ratti}
\affiliation{SENSEable City Laboratory,
Massachusetts Institute of Technology, 77 Massachusetts Avenue,
Cambridge, MA 02139, USA}



\date{\today}

\begin{abstract}
\begin{it}
Recent years have witnessed the development of a large body of algorithms for community detection in complex networks.
Most of them are based upon the optimization of objective functions, among which modularity is the most common, though a number of alternatives have been suggested in the scientific literature.
We present here an effective general search strategy for the optimization of various objective functions for community detection purposes.
When applied to modularity, on both real-world and synthetic networks, our search strategy substantially outperforms the best existing algorithms in terms of final scores of the objective function.
In terms of execution time for modularity optimization this new approach also outperforms most of the alternatives present in literature with the exception of fastest but usually less efficient greedy algorithms.
The networks of up to $30\,000$ nodes can be analyzed in time spans ranging from minutes to a few hours on average workstations, making our approach readily applicable to tasks not limited by strict time constraints but requiring the quality of partitioning to be as high as possible.
Some examples are presented in order to demonstrate how this quality could be affected by even relatively small changes in the modularity score stressing the importance of optimization accuracy.
\end{it}
\end{abstract}

\keywords{Complex networks | Community detection | Network science}

\maketitle

The increasing availability of big data has motivated an enormous general interest in the burgeoning field of network science.
In particular, the broad penetration of digital technologies in different spheres of human life provides substantial sources of data sets which explore the intricacies of manifold aspects of human activity. The topics they cover range from personal relationships among individuals to professional collaborations, from telephone communication to data exchange, from mobility and transportation to economical transactions and interactions in social media.
Analyzing such data sets often leads to the construction of complex networks describing relations among individuals, enterprises, locations, or more abstract entities, such as the buzzwords and hashtags employed in social media; whenever the resulting structures are geographically located, they can then be studied at different scales, including global, countrywide, regional, and local levels.
Furthermore, complex networks can arise from the study of biological phenomena, including neural, metabolic, and genetic interactions.

Community detection is one of the pivotal tools for understanding the underlying structure of complex networks and extracting useful information from them; it has been used in fields as diverse as biology\,\cite{Guimera2005FunctionalCartography}, economics - the World Trade Net is analyzed in\,\cite{PiccardiWorldTradeWeb} - 
human mobility\,\cite{Thiemann2010StructureOfBorders, HossmannHumanMobility, amini2014impact, hawelka2014geo, kang2013exploring}, communications\,\cite{Ratti2010GB, Sobolevsky2013delineating}, and scientific collaborations\,\cite{Palla2005Overlapping}.
Many algorithms were devised in the field of community detection, ranging from straightforward partitioning approaches, such as hierarchical clustering\,\cite{Hastie2001ElementsOfStatisticalLearning} or the Girvan-Newman\,\cite{GN} algorithm, to more sophisticated optimization techniques based on the maximization of various objective functions.

The most widely used objective function for partitioning is modularity\,\cite{newman2004,newman2006}: it relies on comparing the strength of inter- and intra-community connections with a null-model in which edges are randomly re-wired.
In order to obtain partitions yielding optimal values for modularity, researchers have suggested a large number of optimization strategies:
well-known algorithms include the simple greedy agglomerative optimization by Newman\,\cite{NewmanPRE2004} and faster Clauset-Newman-Moore heuristic\,\cite{CNM2004VeryLargeNetworks};
Newman's spectral division method\,\cite{newman2004} and its improvements (which employ an additional Kernighan-Lin-style\,\cite{Kernighan70} step)\,\cite{newman2006};
a similar method by Sun et al.\,\cite{Sun2009}, in which partitions are iteratively refined by considering all possible moves of single nodes to all existing or new communities;
the aggregation technique commonly referred to as Louvain method, extremely fast even on large-scale networks\,\cite{leuven};
simulated annealing\,\cite{simulatedAnnealing,Good2010PerformanceOfModularity};
extremal optimization\,\cite{Duch2005CElegans}; and many others\,\cite{fortunato2010}.
In the last few years, researchers have shown that modularity suffers from certain drawbacks, including a resolution limit\,\cite{Fortunato02012007ResolutionLimit, Good2010PerformanceOfModularity} which prevents it from recognizing smaller communities (a proposed multi-scale workaround which involves modifying the network can be found in\,\cite{Arenas2008Analysis}).

At least three of the several alternative objective functions deserve to be mentioned: description code length, block model likelihood measure, and surprise.
The description code length of a random walk on a network,
upon which the Infomap algorithm\,\cite{Rosvall01052007InformationTheoretic, Infomap} by Rosvall and Bergstrom is based, is an well-known information-theoretical measure, reputed to be among the best available\,\cite{Lancichinetti2009CommunityComparative};
it appears, however, that code length optimization also suffers from a resolution limit, as discussed in\,\cite{BrantingInfoTheoreticCriterion}, where a workaround is proposed.
The second approach is based on the likelihood measure for the stochastic block model, variations of which were suggested in\,\cite{Newman2011Stochastic,Newman2011Efficient,Bickel2009Nonparametric, Decelle2011BlockModel, Decelle2011BlockModelAsymptotics, Yan2012ModelSelection}.
Finally, Surprise\,\cite{Aldecoa2011Deciphering} compares the distribution of inter-community links to that emerging from a random network with the same distribution of nodes per community.
For a detailed, if not up-to-date, review of existing community detection methods, the reader can refer to Ref.\,\cite{fortunato2010}.

A few more strategies for community detection follow:
the replica correlation method introduced in\,\cite{RonhovdeMultiresolution}, which is also an information-based measure;
two recently proposed algorithms, which infer community structures by using generalized Erd\H{o}s Numbers\,\cite{Morrison2012CommFriendship} and by focusing on the statistical significance of communities\,\cite{Lancichinetti2011StatisticallySignificant};
a recent approach for modularity optimization - conformational space annealing\,\cite{LeeCSA} - which delivers acceptable results very quickly, and is scalable to larger networks, as is the modification to the algorithm by Clauset, Newman, and Moore\,\cite{CNM2004VeryLargeNetworks} proposed in\,\cite{Wakita2007MegaScaleSocial}.

A key point in the evaluation of algorithms for community detection is the choice of meaningful benchmarks.
Benchmarks can be roughly divided into two groups. 
In the first, one compares the final scores achieved by different algorithms for the optimization of the same objective function on a variety of networks.
In the second type of benchmark, resulting partitions are checked against imposed or well-known structures in synthetic or real-world networks; this kind of benchmark is fundamental for the evaluation of different partitioning techniques not necessarily based on the optimization of the same objective function.
Other methods to obtain independent evaluations of the reliability of communities found, without relying on the known community structure nor objective function scores, focus~-- among other parameters~-- on recurrence of communities under random walks\,\cite{Delvenne20072010, LeMartelot2011Stability}, and their resilience under perturbations of the network edges\,\cite{Mirshahvalad2012SignificantCommunities}.

In the present work we suggest a novel universal optimization technique for community detection, which we apply to two of the aforementioned objective functions: modularity and description code length.
We also present the results of a two-stages benchmark.
First, we compare the performance of our algorithm, in terms of the resulting values for objective functions, with a host of existing optimization strategies, separately for modularity and description code length; we show in this way that we consistently provide the best modularity scores, and results on par with Infomap when optimizing description code length.
Next, by employing in each case the best available algorithm, we compare the performances of modularity and description code length as objective functions in reconstructing underlying structures on a large set of synthetic networks, as well as the known structures on a set of real-world networks.

\vspace{1ex}\noindent\textbf{1. The algorithm}

\noindent
The vast majority of search strategies take one of the following steps to evolve starting partitions: merging two communities, splitting a community into two, moving nodes between two distinct communities.
The suggested algorithm involves all three possibilities. After selecting an initial partition made of a single community,
the following steps are iterated as long as any gain in terms of the objective function score can be obtained: (\textbf{1}) for each source community, the best possible redistribution of every source nodes into each destination community (either existing or new) is calculated; this also allows for the possibility that the source community entirely merges with the destination; (\textbf{2}) the best merger/split/recombination is performed. As the proposed technique combines all three possible types of steps, in the following we'll refer to it as Combo.

The fulcrum of the algorithm is the choice of the best recombination of vertices between two communities, as splits and mergers are particular cases of this operation:
for each pair of source and (possibly empty) destination communities, we perform a shift of all the vertices fashioned after Kernighan and Lin's algorithm\,\cite{Kernighan70}.
Specifically, we recombine the two communities starting from several initial configurations, which include \textbf{(a)} the original communities, \textbf{(b)} the case in which the whole source community is moved to the destination, \textbf{(c)} a few intermediate mergers, in which a random subset of the source community is shifted to the destination.
For each starting configuration, we iterate a series of Kernighan-Lin shifts until no further improvement is possible; each is performed by: \textbf{(1)} initializing a list of available nodes to include all the nodes from the original source community; \textbf{(2)} iterating the following steps until list is empty: \textbf{(a)} find the node $i$ in the list for which switching community entails the largest gain or the minimum loss (if no gains are available); \textbf{(b)} switch $i$ to the other community, remove $i$ from the list of available nodes, and save the intermediate result.
After a series of Kernighan-Lin improvements has been completed for each of the starting configurations, we select the intermediate result which yields the best score in terms of objective function.
See Algorithm~\ref{alg:combo_pseudocode} for schematic pseudocode of Combo\footnote[1]{The C++ implementation of the Combo algorithm used in this paper could be downloaded from \url{http://senseable.mit.edu/community_detection/combo.zip}}.

{
\begin{algorithm}[bth]
\caption{Combo}\label{alg:combo_pseudocode}

\fontsize{11}{10}\selectfont
\setlength{\leftskip}{0.5em}
\SetInd{1em}{1em}

\SetKwInOut{Input}{input}\SetKwInOut{Output}{output}
\SetKwProg{proc}{Procedure}{}{}
\SetKwFunction{reCalc}{ReCalculateGain}
\SetKwFunction{KernighanLin}{PerformKernighanLinShifts}
\SetKwFunction{move}{PerformMove}
\SetKwFunction{bestGain}{BestGain}
\Input{A network $net$ containing $n$ nodes, initial partition $initial\_communities$ (by default initially all nodes in one community), the maximal number of communities $max\_communities$ ($infinity$ by default)}
\Output{A partition of the network into $communities$}
\BlankLine
  \emph{Initialize variables for storing partitions and their gains}\;
  
  \For(\tcp*[h]{$dest$ may be empty community}){each pair $(origin, dest)$ of communities}{
    \tcp{Calculate best gain from moving nodes from origin to dest}
    \reCalc{origin, dest}\;
  }
  
  \While{\bestGain{} $> THRESHOLD$}{
    \move{best\_origin, best\_dest, best\_partition}\;
	
	\tcp{Update gains for changed communities}
	\For{each community $i$}{
      \reCalc{best\_origin, i}; \reCalc{i, best\_origin}\;
      \reCalc{best\_dest, i}; \reCalc{i, best\_dest}\;
    }
  }
  \BlankLine
  \proc{\move{origin, dest, partition}}{
    \emph{Move nodes from $origin$ to $dest$ according to $partition$}\;
  }
  \BlankLine
  \proc{\bestGain{}}{
    \emph{Select from remembered partitions one with the best gain}\;
    \emph{Return this $gain$ and corresponding $best\_origin$, $best\_dest$ and $best\_partition$}\;
  }
  \BlankLine
  \proc{\reCalc{origin, dest}}{
    \If{$dest$ is new community \textbf{and} we already have $max\_communities$}{
      \KwRet\;
    }
    \emph{Define and initialize number\_of\_tries}\;
    \For{$tryI \leftarrow 1$ \KwTo $number\_of\_tries$}{
      \ForEach{vertex $v$ from $origin$ community}{
        \emph{move $v$ to $dest$ or leave in $origin$ with equal probability}\;
      }
      \emph{Calculate new gain, assign zero to previous gain}\;
      \While{new gain $>$ previous gain}{
        \KernighanLin{origin, dest}\;
      }
      \If{achieved gain is greater then current maximum}{
        \emph{Remember current partition and gain}\;
      }
    }
  }
  \BlankLine
  \proc{\KernighanLin{origin, dest}}{
    \emph{Calculate gains from moving each node to opposite community}\;
    \For{$i \leftarrow 1$ \KwTo size of origin community}{
      \emph{Perform temporary movement that produces maximal gain}\;
      \emph{Remember current gain and moved node}\;
      \emph{Recalculate all gains}\;
    }
    \emph{Retrieve the movements leading to a maximal gain among intermediately calculated and perform them}\;
  }
\end{algorithm}
}

It's also worth mentioning that the inclusion of random initial configurations is usually essential to the algorithm performance.
The experiments reported in Supplementary Material\,\cite{SI} on Fig.\,S5 show that on average considering random configurations increases the resulting modularity score by $2\%$, which could sometimes correspond to quite a considerable partitioning improvement.
As we can see in table\,\ref{tab:lowModDeltaLargeNMI} even much smaller changes to modularity score result in significant variations in the partitioning.
Also Fig.\,S6 from Supplementary Material\,\cite{SI} shows that despite this randomness results of Combo are very stable (varying in bounds of $0.1\%$).
However processing random configurations also takes time~-- without them the algorithm appears to be on overage $4.2$ times faster, which makes it possible to suggest this simplified version of the algorithm for the applications when execution time is more crucial.
At the same time, replacing such random configurations with partitioning produced via other methods, e.g. spectral division, makes the algorithm more prone to being captured by local maxima.

\begin{figure*}[hbtp]
\centering
\includegraphics[width=.9\textwidth]{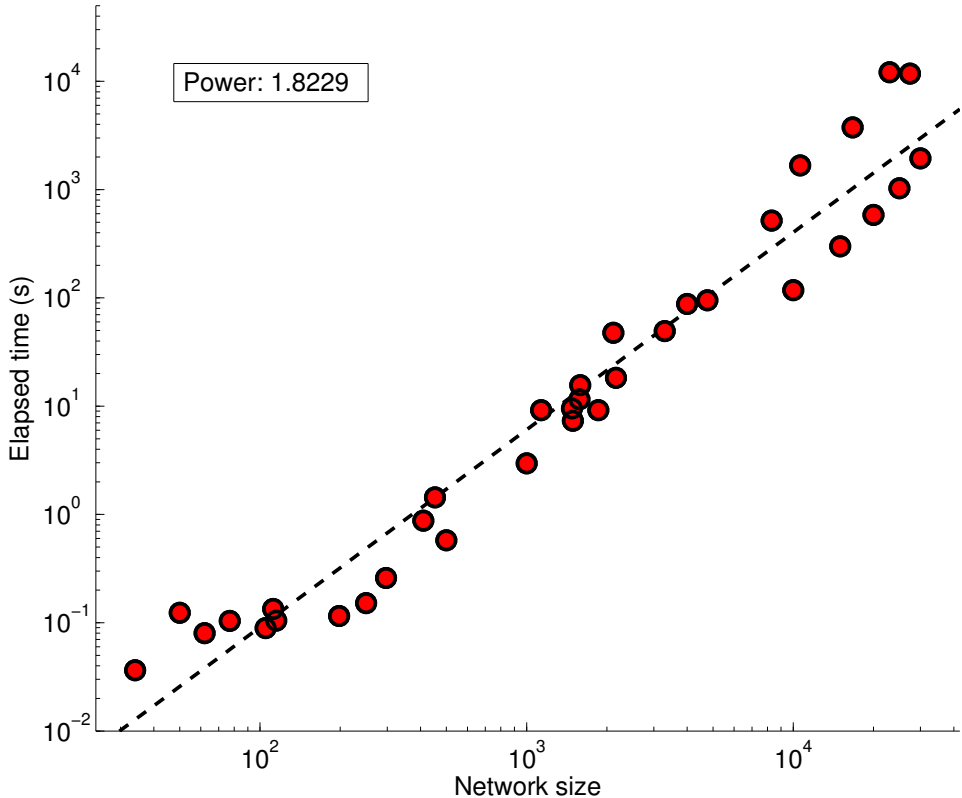}
\caption{\label{fig:ComboTimeVsSize}(Color online) Dependence of Combo execution time on the network size (for all the benchmark networks described below) showing a power law relation.}
\end{figure*}

Experimental tests show a striking regularity in the dependence of Combo execution time on the number of nodes of the network;
Fig.\,\ref{fig:ComboTimeVsSize} demonstrates that this behaviour is close to a power law with exponent $1.8$.
As one can see from the figure Combo can deal with networks of up
to $30\,000$ nodes in time of up to a few hours (on iMac machine with Core i7 3.1\,GHz CPU and 16\,GB memory).
However memory availability is a bottle-neck of the current implementation and for the bigger networks the code slows down even more whenever it starts using computer's virtual memory.

As the sequence of operations in Combo is strongly dependent on the specific network, sharp evaluations of its computational complexity are difficult to obtain; the regularity of the dependence observed in Fig.\,\ref{fig:ComboTimeVsSize} - however - hints at some robust mechanism acting under the hood. In the Supplementary Material\,\cite{SI}, we justify an upper bound to the execution time of ${\cal O}\left(N^2\log\left({\cal C}\right)\right)$, where $N$ is the number of nodes, and ${\cal C}$ the number of communities in the network.

\begin{figure*}[t!]
\centering
\includegraphics[width=1.\textwidth]{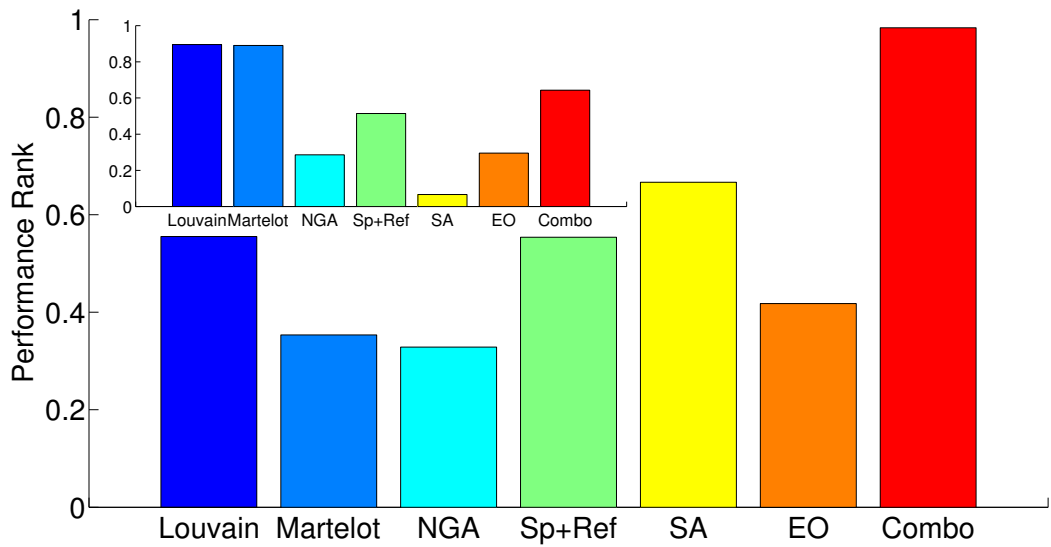}
\caption{\label{fig:Ranks}(Color online) Average normalized performance rank of each algorithm in terms of partitioning quality (main plot) and speed (subplot): values ranging from 0 (worst performance) to 1 (best) are attributed to each algorithm, and their average computed.}
\end{figure*}

\begin{figure*}[tbhp]
\centering
\includegraphics[width=1.\textwidth]{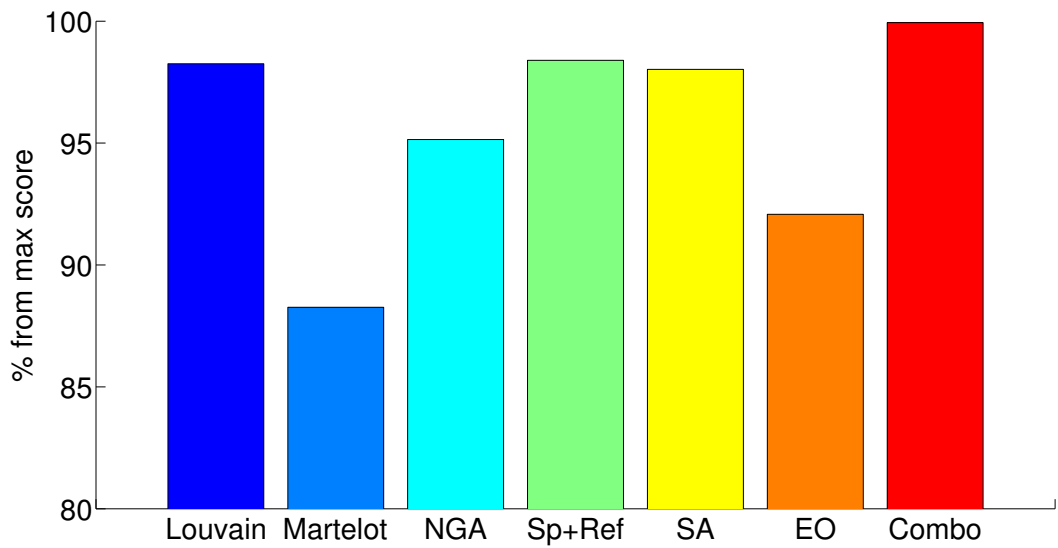}
\caption{\label{fig:PercentsFromMax}(Color online) Performance of algorithms as average percent of their resulting modularity score to the maximum, achieved by the best algorithm.}
\end{figure*}

\begin{figure*}[t!]
\centering
\includegraphics[width=\textwidth]{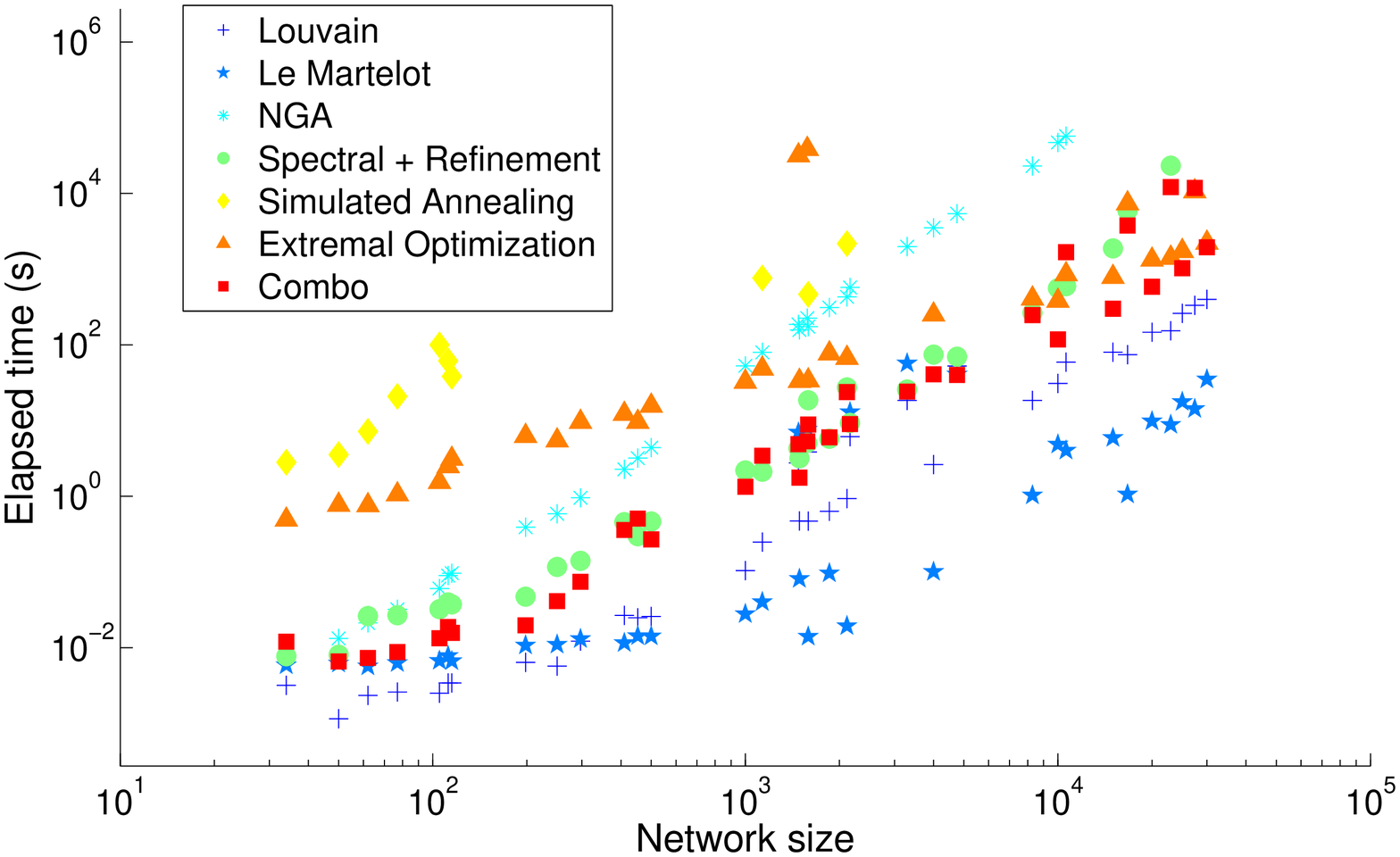}
\caption{\label{fig:execTimes}(Colored online) Execution times by network size and algorithm.}
\end{figure*}

\vspace{1ex}\noindent\textbf{2. Modularity optimization benchmarks}

\noindent
We first evaluated the performance of Combo for modularity optimization.
We selected six algorithms for the comparison:
\textbf{a)} Louvain method\,\cite{leuven};
\textbf{b)} Le Martelot\,\cite{LeMartelot2011Stability};
\textbf{c)} Newman's greedy algorithm (NGA)\,\cite{NewmanPRE2004};
\textbf{d)} Newman's spectral algorithm with refinement\,\cite{newman2006};
\textbf{e)} Simulated annealing\,\cite{simulatedAnnealing};
\textbf{f)} Extremal optimization\,\cite{Duch2005CElegans}.
The set of algorithms we have chosen offers a good sample of the current state of the art.
Simulated annealing is reputed to be capable of getting very close to real maxima, and extremal optimization offers a good tradeoff between speed and performance\,\cite{fortunato2010,Liu20102300,Aldecoa2013SurpriseMaximization}; they resulted the best-performing algorithms in at least one benchmark\,\cite{Danon2005}. The recursive Louvain method is fast and relatively effective\,\cite{Lancichinetti2009CommunityComparative} and has therefore been applied in various real-world network analyses\,\cite{Meunier2009,Zhang2010}.
Newman's greedy algorithm and Spectral Algorithms
can be considered classical approaches, since they were suggested right after modularity was introduced about 10 years ago, and were therefore used in a number of previous benchmarks\,\cite{fortunato2010,Aldecoa2013SurpriseMaximization,leuven,Lancichinetti2009CommunityComparative}.
The technique by Le Martelot is a more recent approach, for which a benchmark already exists\,\cite{lemartelot2012}.

We ran each algorithm on three sets of networks: \textbf{(1)}~widely available data sets found in literature; \textbf{(2)}~five graphs - obtained from NDA-protected telecom data - in which the weight of each edge corresponds to the total duration of telephone calls between two locations; \textbf{(3)}~ten synthetic networks generated using the Lancichinetti-Fortunato-Radicchi approach\,\cite{LFR,LFR2}.
Detailed descriptions and references can be found in the Supplementary Material\,\cite{SI}.

As a measure of the comparative quality of partitioning, we computed the average rank of each algorithm over all the networks on which it has been tested.
When multiple algorithms yielded the same modularity, we equated their rank to the best among them (1 for the highest modularity score).
For ranks based on execution time we scored zero all those algorithms that didn't converge within 12 hours.

As summarized in Fig.\,\ref{fig:Ranks} and Fig.\,\ref{fig:PercentsFromMax}, Combo significantly outperforms other algorithms, with an average rank score of $0.98$; the next best placements are Simulated Annealing ($0.67$), Louvain ($0.55$), and Spectral method ($0.51$);
other algorithms show considerably less consistent outcomes.
Fig.\,\ref{fig:execTimes} shows that Combo is not as fast as the greedy aggregation algorithms (Louvain, Le Martelot), but faster than other algorithms, both complex, such as Simulated Annealing, and simple, as NGA (for which we are however using a Matlab implementation).
In the worst cases (usually when the resulting number of communities is big enough), Combo finalizes computation in a matter of hours for networks of thousands to tens of thousands of nodes.
That is why in cases where the network is big enough and the computational time is crucial, while the resulting partitioning quality is not, using faster approaches might be the better choice.

Often, however, the reliability of the final community structure is of paramount importance: in such cases, we'll want to aim at the highest possible value of the objective function, as even small differences in the resulting modularity score can translate into macroscopic variations in the quality of partitioning.
In the next section, we show that a variation as small as $0.5\%$ can have a sizable impact on the community structure of a network, and Fig.\,\ref{fig:PercentsFromMax} demonstrates that Combo outperforms its nearest rivals by around 2\% on average in terms of achieved modularity score.
While at the moment it's impossible to guarantee that an achieved partition is a global maximum, we can assume that choosing the one sporting the highest score is the best option.

\vspace{1ex}\noindent\textbf{3. Importance of Precision: The Effect of Small Changes in Modularity Values on Partitions}

Here in order to stress the importance of looking for even the minor gains in the modularity score, we would like to show that relatively small changes in this partition quality function can be reflected by macroscopic variation of the communities involved.
To illustrate this point, at first, we compared the partition with the highest modularity score of ten first networks (incidentally for all ten networks it is the one obtained by using Combo) from our modularity benchmark (their descriptions can be found in Supplementary Material\,\cite{SI}) with the partitioning obtained by Louvain method being one of the closest competitors.
As shown in table\,\ref{tab:lowModDeltaLargeNMI}, differences in modularity score that one might consider to be relatively low can correspond to sizeable variations of partition.
In order to quantify that difference we used normalized mutual information (NMI)\,\cite{Danon2005} (introduced in detail in the Supplementary Material\,\cite{SI}).
It is scaled from $0$ to $1$ and the more similar partitions are the higher NMI they have, for identical partitions NMI equals to $1$.
We see that quite often difference in the modularity score less than $0.01$ or even $0.001$ which one might perhaps consider to be the minor deviation at the first glance, could actually result in substantial variations of the corresponding community structure with the corresponding NMI similarity values sometimes as low as $0.6-0.7$.

\begin{table*}[htb]
\caption{\label{tab:lowModDeltaLargeNMI} Difference in the modularity score and corresponding NMI similarity between best and alternative partitioning produced by different algorithm.}
\begin{center}
\begin{tabular}{| c | c c | c | c |}
\hline
\multirow{2}{*}{Network} & \multicolumn{2}{c|}{Modularity score} & \multirow{2}{*}{Deviation} & \multirow{2}{*}{NMI} \\
                         & Best         &            Alternative &                            &  \\
\hline
 1 & 0.419790 & 0.418803 & 0.000987 & 0.923345\\
 2 & 0.526799 & 0.518828 & 0.007971 & 0.732029\\
 3 & 0.566688 & 0.565416 & 0.001272 & 0.924726\\
 4 & 0.527237 & 0.498632 & 0.028605 & 0.784013\\
 5 & 0.310580 & 0.290605 & 0.019975 & 0.553769\\
 6 & 0.605445 & 0.602082 & 0.003363 & 0.919872\\
 7 & 0.507642 & 0.493481 & 0.014161 & 0.741351\\
 8 & 0.432456 & 0.432057 & 0.000399 & 0.651622\\
 9 & 0.955014 & 0.954893 & 0.000121 & 0.971705\\
10 & 0.850947 & 0.846159 & 0.004788 & 0.816490\\
\hline
\end{tabular}
\end{center}
\end{table*}

Another important question of course is whether those noticeable changes in community structure sometimes coming along with the small gains in the modularity scores one could achieve by using the higher performance algorithm, actually improve the partitioning quality in a certain sense.
This is a complex question laying mostly beyond the scope of the current article as in fact it requires one to understand to which extent the modularity score itself could be trusted as the partitioning quality function.
There is an ongoing debate in the literature about advantages and limitations of the modularity optimization approach including the modularity resolution limit\,\cite{Fortunato02012007ResolutionLimit, Good2010PerformanceOfModularity}.
Also the question of what to take for a partitioning quality is not always obvious~-- even if for some of the real-world networks we possess a knowledge of their actual underlying community structure there is no guarantee it would be indeed optimal in any theoretical sense including modularity score optimization.
But just as a simple illustration to that question we introduce a second experiment where we used networks generated by Lancichinecchi-Fortunato-Radicchi's method\,\cite{LFR, LFR2} having a pretty much straightforward imposed community structure.
For each of the networks we compared two partitions obtained by Combo and Louvain method with this original community structure based on which the network was created.
Table\,\ref{tab:ComboLouvainLFRnmi} shows that while the results of Combo providing the better modularity score appear to be $99-100\%$ similar to the original community structure, the results of the other method that might seem to be just slightly worse in terms of modularity, already demonstrate a much less convincing match~-- usually around $95-97\%$ but sometimes down to $70\%$ or even $15\%$ in terms of NMI.
And better modularity score always comes together with the better NMI.

\begin{table*}[hbt]
\caption{\label{tab:ComboLouvainLFRnmi}NMI similarity to the original network structure and corresponding modularity scores for partitioning of LFR synthetic networks produced by different algorithms.}
\begin{center}
\begin{tabular}{| c | c c | c | c c |}
\hline
Network & \multicolumn{2}{c|}{Modularity score} & \multirow{2}{*}{Deviation} & \multicolumn{2}{c|}{NMI} \\
size    & Best         &            Alternative &                            &  Best   &    Alternative \\
\hline
 1000 & 0.376667 & 0.342281 & 0.034386 & 0.989395 & 0.705298\\
 2000 & 0.339416 & 0.243512 & 0.095904 & 0.998217 & 0.158417\\
 3000 & 0.569376 & 0.556105 & 0.013271 & 1.000000 & 0.969851\\
 4000 & 0.570596 & 0.563286 & 0.007310 & 0.997617 & 0.975451\\
 5000 & 0.616145 & 0.609881 & 0.006264 & 0.996095 & 0.976030\\
 6000 & 0.571150 & 0.556786 & 0.014364 & 0.996035 & 0.954530\\
 7000 & 0.565559 & 0.549285 & 0.016274 & 0.996824 & 0.944399\\
 8000 & 0.614574 & 0.608583 & 0.005991 & 0.991510 & 0.968284\\
 9000 & 0.575881 & 0.566198 & 0.009683 & 1.000000 & 0.961747\\
10000 & 0.605243 & 0.581807 & 0.023436 & 0.996522 & 0.943804\\
\hline
\end{tabular}
\end{center}
\end{table*}

\begin{figure*}[htb]
\centering
\includegraphics[clip,trim={10, 60, 80, 60},width=.60\textwidth]{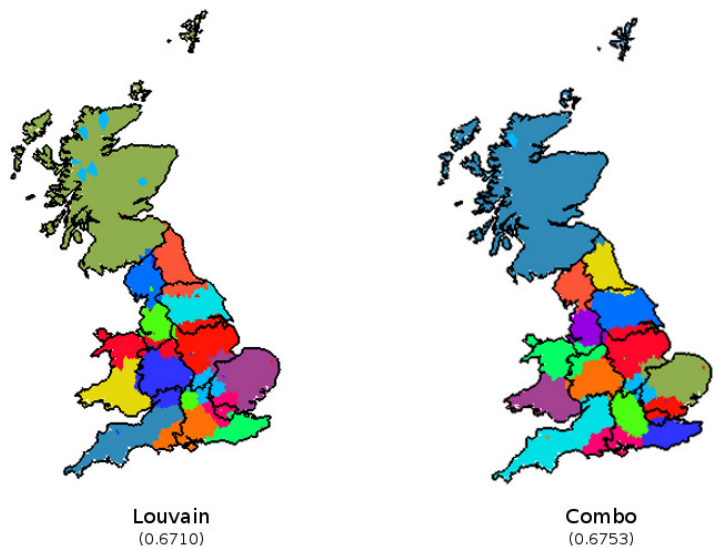}
\caption{\label{fig:UKComboVsLouvainNCALLS}(Colored online) Partitioning of a network based on the number of all calls entertained between each pair of locations. A minimal variation in modularity (less than 1 percent) can turn into a sizable difference in partitioning. Here the Combo results show cleaner geographical separation of communities and are substantially more similar to official administrative divisions with modularity equal to $0.6753$ and NMI $=0.804$, compared to modularity $=0.6710$ and NMI $=0.703$ for Louvain.}
\end{figure*}

Just to give a visual example of how the partitioning changes corresponding to the minor modularity improvement could look like we show two different partitions for the United Kingdom telephone network studied in\,\cite{Ratti2010GB, Sobolevsky2013delineating} in which link weights represent the number of telephone calls between locations: one obtained with Combo, the other with the Louvain method (Fig.\,\ref{fig:UKComboVsLouvainNCALLS}). Although the modularity gain is only $0.0043$, which at a first glance may suggest that the quality of two partitioning is actually comparable, a number of macroscopic differences are visible.
Slightly higher modularity score also translates into a lower level of noise in the spatial structure of the resulting communities and a better agreement with the official administrative divisions of Great Britain being quantified by NMI similarity measure~-- $0.804$ against $0.703$.

\vspace{1ex}\noindent\textbf{4. Minimum description code length benchmarks}

In our second benchmark, we use the combo algorithm to optimize description code length compression, and compare the results to those obtained using the original Infomap implementation by Rosvall and Bergstrom\,\cite{Rosvall01052007InformationTheoretic, Infomap}.
Because of longer execution time of Combo for code length, we ran the comparison on the set of networks of size up to $8000$ from the previous benchmark.
Since Infomap is a greedy algorithm and results are dependent on a random seed, we ran it 10 times for each network and picked the best result.

Unlike for modularity, final values for code length are very close, with a single network in which their difference is about $5\%$, and less than $3\%$ in all other cases; Combo yields a better code length in 8 networks, Infomap in 9, the results being the same in all other cases. Detailed results are reported in the Supplementary Material\,\cite{SI}.
Combo thus results a valid alternative and an ideal complement to Infomap, as in several cases it's proved capable of finding better solutions.

The analysis above proves that Combo is efficient in terms of optimization of both objective functions~-- modularity and code length.
Now having such a high-performance universal optimization technique opens a new research opportunity worth additional consideration.
Some attempts at comparing multiple partitioning algorithms dealing with different objective functions are already present in literature\,\cite{Laarhoven2013}.
However if done using different optimization techniques it is not possible to clearly judge whether higher performance of a certain approach is due to the objective function relevance or just the optimization technique performance.
As Combo efficiently yields near-optimal results for both modularity and code length, we can now for the first time fairly compare modularity and code length as community detection objective functions.
As a simple initial criteria for such a comparison we consider the ability of reproducing the existing pre-imposed community structure in synthetic networks.
Results are presented in the Supplementary Material (see section Modularity vs. Description Code Length Comparison)\,\cite{SI}.
Overall we found that modularity yields more reliable community reconstruction in more complex cases as the level of noise increases.
Also code length performs surprisingly poorly for smaller networks, while for bigger networks with relatively low level of noise its performance already exceeds the one of modularity.
Based on that, one could recommend using modularity for discovering community structure in networks with weaker clustering effect, while code length might be a better choice for larger networks with relatively strong communities.

\vspace{1ex}\noindent\textbf{5. Conclusions}

We have presented Combo, an optimization algorithm for community detection capable of handling various objective functions, and we analyzed its performance with the two most popular partitioning quality measures: modularity and description code length.
With regard to modularity, Combo consistently outperforms all the other algorithms with which we have compared it, including the current state of the art.
For what concerns the code length optimization, Combo provides results on par with those of Infomap, which is the defining algorithm for this objective function.

The current implementation of Combo however has limitations in terms of maximal network size it is able to handle within a reasonable time: due to memory constrains its current applicability limit is around $30\,000$ nodes on modern workstations.
Running times are usually longer compared to the fastest greedy algorithms, but often considerably shorter than for other highly efficient optimization techniques: networks whose size is close to the above threshold can be handled within a few hours, while smaller networks of several thousand nodes only require minutes.
Combo is thus an optimal choice when the quality of the resulting partition is of paramount importance, while the network is not too big and running time is not strictly constrained.

Combo as an optimization technique is flexible, in that it can be adapted to many other objective functions; possible extensions might be stochastic block model likelihood\,\cite{newman2013community} and surprise\,\cite{Aldecoa2011Deciphering}.
Additional advantages include the possibility of limiting the number of resulting communities (e.g. to obtain the optimal bi-partitioning of a network) and the algorithm applicability to further fine-tuning of results previously obtained using other algorithms.

Finally, by studying how well the most efficient optimization techniques for modularity and code length reproduce the known underlying community structure of the networks, we have provided as fair as possible a comparison between the two objective functions.

\vspace{1ex}\noindent\textbf{Acknowledgements}

The authors wish to thank Orange and British Telecom for providing some of the datasets used in this study.
We further thank National Science Foundation, the AT\&T Foundation, the MIT SMART program, the Center for Complex Engineering Systems at KACST and MIT, Volkswagen, BBVA, The Coca Cola Company, Ericsson, Expo 2015, Ferrovial, the Regional Municipality of Wood Buffalo, AIT, and all the members of the MIT Senseable City Lab Consortium for supporting the research.

The authors also want to thank Paolo Santi for helpful discussions, and Kael Greco for his help with graphics.

\bibliographystyle{apsrev}
\bibliography{combo}

\end{document}